\documentclass[twocolumn,showpacs,preprintnumbers,amsmath,amssymb]{revtex4}
\usepackage{graphicx}

\begin{document}

\title{Degenerate approach to the mean field Bose- Hubbard Hamiltonian}

\author{A.M. Belemuk and V.N. Ryzhov}
\address{Institute for High Pressure Physics, Russian Academy
of Sciences, Troitsk 142190, Moscow Region, Russia}


\begin{abstract}
A degenerate variant of mean field perturbation theory for the
on-site Bose-Hubbard Hamiltonian is presented. We split the
perturbation into two terms and perform exact diagonalization in
the two-dimensional subspace corresponding to the degenerate
states. The final relations for the second order ground state
energy and first order wave function do not contain singularities
at integer values of the chemical potentials. The resulting
equation for the phase boundary between superfluid and Mott states
coincides with the prediction based on the conventional mean field
perturbation approach.
\end{abstract}

\pacs{03.75.Lm,03.75.Kk,67.57.Fg}

\maketitle

The bosonic Hubbard model has been the subject of intense
theoretical interest \cite{Zwerger03, Lewenstein07, Weichman08}.
It describes quantum phase transition between the superfluid and
Mott phases of ultracold bosonic atoms in an optical lattice,
first realized in experiment by {\it Greiner et al.}
\cite{Greiner02}. A qualitatively correct phase diagram of the
model at zero temperature can be captured by a simple mean field
theory \cite{Fisher89, Rokhsar91, Krauth92, Jaksch98, Sheshadri93,
Stoof01}. This theory predicts superfluid phase for non-integer
fillings and a transition to an insulating Mott phase for integer
fillings. This findings were confirmed with numerical calculations
(quantum Monte Carlo simulation) \cite{Kashurnikov02, Alet04,
Capogrosso07}, variational approach \cite{Capello07}, cluster
Gutzwiller method \cite{Luhmann13}, and more elaborate analytical
methods, including strong-coupling perturbation theory
\cite{Freericks96}, slave-boson theory \cite{Ziegler93, Fresard94,
Stoof03}, effective action approach \cite{Sengupta05, Santos09,
Bradlyn09}, random phase approximation \cite{Ohashi06, Konabe06,
Menotti08}, bosonic dynamical mean-field theory \cite{Kauch12},
multisite mean-field theory \cite{Buonsante04, Buonsante05,
McIntosh12}, higher-order series expansions \cite{Eckardt09,
Teichmann09}.

In this paper, we present an alternative estimation of the
zero-temperature ground state wave function $|\psi_{gs} \rangle$
and energy $E_{gs}$, which takes into account the degenerate
character of the ground state of $H_0$ at integer values of the
chemical potential $\mu$. The resulting formulas for $|\psi_{gs}
\rangle$ and $E_{gs}$ are free from singularities which otherwise
are present at integer values of the chemical potentials in the
conventional perturbative mean field approach \cite{Stoof01}.

For this we perform first the diagonalization  of the on-site
mean-field Hamiltonian in the basis of the two lowest states at a
given $\mu$, and the account for the rest of the perturbation  by
a conventional technique. The properties of the Mott phase and the
superfluid phase in a Hilbert space restricted to the number-basis
states $|n \rangle$ and $|n+ 1 \rangle$ (the corresponding
Gutzwiller state is $\prod_i |\psi \rangle_i$, with $|\psi
\rangle= f_n |n \rangle+ f_{n+1} |n+1 \rangle$) was considered in
Refs.  \cite{Barankov07, Sun09}.

In the framework of mean field approach the hopping term between
cites $i$ and $j$ is decoupled as, $b^{\dagger}_i b_j \simeq
\langle b^{\dagger}_i \rangle b_j+ b^{\dagger}_i \langle b_j
\rangle- \langle b^{\dagger}_i \rangle \langle b_j \rangle$ and
the Bose- Hubbard (BH) Hamiltonian is reduced to on-site
Hamiltonian $H= H_0+ H_t$,
\begin{gather}
H_0= -\mu_0 n + \frac{U_0}{2} n(n-1), \\ H_t= -t_0\nu(\psi^* b+
\psi b^{\dagger})+ t_0\nu |\psi|^2,
\end{gather}
where $U_0$ is a repulsive on-site boson- boson interaction, $H_t$
is the mean field kinetic energy, $t_0$ is the nearest-neighbor
hopping amplitude, $\nu$ is the number of nearest neighbors. The
symbols $b$ and $b^{\dagger}$ denote destruction and creation
operators for bosons at a lattice site, $\psi= \langle b \rangle$
is the order parameter, $n= b^{\dagger}b$ is the boson number
operator. The chemical potential $\mu_0$ controls the average
number of bosons.

It is convenient to rewrite $H$ in dimensionless units $\mu_0/U_0=
\mu$, $t_0\nu/U_0= t$ and $H/U_0= h$. In these units
\begin{gather} \label{ham}
h= h_0+ h_t+ t|\psi|^2, \\
h_0= -\mu n + \frac{1}{2} n(n-1), \quad h_t= -t(\psi^* b+ \psi
b^{\dagger})
\end{gather}

Eigenfunctions of $h_0$ are the number states $|n \rangle$ and
eigenvalues are $\varepsilon_n= n^2/2- (\mu+ 1/2)n$, $n= 0, 1, 2,
\dots$. The ground state of $h_0$, $|\psi_{gs}^{(0)} \rangle= |n
\rangle$, $E^{(0)}_{gs}= \varepsilon_n$, corresponds to the number
of bosons $n$ if the chemical potential $\mu$ is in the range $n-
1 < \mu < n$. At integer values of $\mu$ the ground state is
two-fold degenerate: $\varepsilon_n= \varepsilon_{n+1}$,
$|\psi_{gs}^{(0)} \rangle= \{ |n \rangle, |n+1 \rangle \}$ at
$\mu= n$, and $\varepsilon_{n-1}= \varepsilon_{n}$,
$|\psi_{gs}^{(0)} \rangle= \{ |n-1 \rangle, |n \rangle \}$ at
$\mu= n- 1$.

The usual practice to handle the perturbation $h_t$ is numerical
diagonalization of $h$ in the subspace spanned by the vectors $|n
\rangle$, $n= 0, 1, \cdots, n_{max}$. This approach essentially
equivalent to the formulation based on the Gutzwiller wave
function $|\psi \rangle= \prod_{i} \sum_{n= 0}^{n_{max}} f^{(i)}_n
|n \rangle_i$, see Refs. \cite{Jaksch98, Lewenstein07, Trefzger08,
Luhmann13}.

Another method is a perturbative treatment of the hopping term.
Corrections to the ground state wave function and energy in the
framework of standard non-degenerate perturbation theory
\cite{Stoof01, Sachdev99} are
\begin{multline} \label{wf1}
|\psi_{gs} \rangle= |\psi_{gs}^{(0)} \rangle+ |\psi_{gs}^{(1)}
\rangle= \\
= |n \rangle + \left[ \frac{(-t) \psi^* \sqrt{n}}{n-1- \mu} |n-1
\rangle+ \frac{(-t) \psi \sqrt{n+1}}{\mu-n} |n+1 \rangle\right],
\end{multline}
\begin{gather}
E_{gs}= \varepsilon_{n}+ t|\psi|^2 \Bigl( 1- t \chi_0(\mu, n)
\Bigr), \label{gs1} \\
\quad \chi_0(\mu, n)= \left[\frac{n+ 1}{n- \mu}+ \frac{n}{\mu- n+
1}\right], \label{chi1}
\end{gather}
where the value of the chemical potential is considered to be in
the interval $n-1 < \mu < n$.

The zero-temperature phase boundary between the Mott state,
$|\psi|= 0$, and the superfluid state, $|\psi|> 0$, corresponds to
the vanishing of the coefficient of the $|\psi|^2$-term of the
expansion of the energy functional $E_{gs}(\psi)$, which gives the
equation $1= t \chi_0(\mu, n)$. The resulting line of critical
values of $t$ as a function of $\mu$ is
\begin{equation} \label{bndr1}
t(\mu, n)= \frac{1}{\chi_0(\mu, n)}= \frac{(n- \mu)(\mu- n +1)}{1+
\mu},
\end{equation}
Although expression for $\chi_0(\mu, n)$, Eq. \eqref{chi1}, is not
defined close to the integer values of $\mu$, the quantity
$1/\chi_0(\mu, n)$ well behaves at integer $\mu$ and, as a result,
the boundary curve, Eq. \eqref{bndr1}, safely includes $\mu= n-1$
and $\mu= n$.

The same boundary equation appears when one considers
self-consistency equation for the order parameter, $\psi= \langle
\psi_{gs} |b |\psi_{gs} \rangle$. First order correction to the
wave function enables to find the first term (linear in $\psi$) in
the expansion $\langle \psi_{gs} |b |\psi_{gs} \rangle= t\psi
\chi_0+ \cdots$, where higher order terms in $\psi$ will come out
if one takes into account next order corrections to the wave
function. The critical boundary corresponds to the vanishing of
the coefficient of the term linear in $\psi$, which amounts to
$\psi= t\psi \chi_0$, and one obtains the same boundary equation.

By construction the standard mean-field scheme cannot describe the
degenerate character of the ground state in the two extremes $\mu=
n-1$ and $\mu= n$ since it is based on the consideration of
interval $n-1 < \mu < n$ and implicitly involves, at $t= 0$,
$|\psi_{gs}\rangle= |n \rangle$ as the reference state. The
corresponding Eqs. \eqref{wf1}, \eqref{gs1} and \eqref{chi1} have
singularities at $\mu= n-1, n$.

Below we show how to change the perturbation expansion to account
for the degenerate case with integers values of $\mu$. This
approach let us identify the phase boundary of the insulating
lobes with no divergence in resulting expressions. Though the
resulting equation for the critical boundary will be the same as
Eq. \eqref{bndr1} this approach provide us information on the
behavior of the order parameter in the superfluid phase.

The basic idea is to make first diagonalization of the
perturbation in two dimensional subspace spanned by vectors ${\cal
P}^2_n= \{ |n \rangle, |n+1 \rangle \}$. It is supposed that the
value of $\mu$ should lie in certain interval around $\mu= n$ so
that $\varepsilon_n$ and $\varepsilon_{n+1}$ are the only two
lowest energies of $h_0$. Below it is convenient to set $\mu= n+
\delta$. Then at $1/2 \leqslant \delta < 1$ the lowest energies
are $\varepsilon_{n+1} < \varepsilon_{n+2} \leqslant
\varepsilon_{n}$, while at $ -1 < \delta \leqslant -1/2 $ the
lowest energies are $\varepsilon_{n} < \varepsilon_{n-1} \leqslant
\varepsilon_{n+1}$. Therefore, the value of $\delta$ is supposed
to lie in the segment $-1/2< \delta < 1/2$. At integer $\mu= n$
this approach accounts for the degenerate level with
$\varepsilon_n= \varepsilon_{n+1}$.

We introduce a projection operator $P$ onto the subspace ${\cal
P}^2_n$ and its orthogonal completion $P^{\perp}$
\begin{equation}
P= |n \rangle \langle n| + |n+1 \rangle \langle n+1|, \quad
P^{\perp}= 1- P
\end{equation}
and rewrite the perturbation as
\begin{gather} \label{ht}
h_t= h'_t+ h''_t, \\
h'_t= P h_t P, \quad h''_t= Ph_tP^{\perp}+ P^{\perp}h_tP+
P^{\perp}h_tP^{\perp}
\end{gather}
The term $h'_t$ we include into $\widetilde h_0= h_0+ h'_t$ and
the term  $h''_t$ we shall treat as a new perturbation. Upon
performing the exact diagonalization of $\widetilde h_0$ in the
two-dimensional subspace ${\cal P}^2_n$,
\begin{equation} \label{h0t}
\widetilde h_0=
\begin{pmatrix}
\varepsilon_{n}   & -t\psi^* \sqrt{n+1} \\
-t\psi \sqrt{n+1} & \varepsilon_{n+1}
\end{pmatrix},
\end{equation}
one obtains  two new zero-order wave functions $|\psi^{(0)}_a
\rangle$ (for lower level) and $|\psi^{(0)}_b \rangle$ (for upper
level)
\begin{gather}
|\psi^{(0)}_{a} \rangle= C_1 |n \rangle+ C_2 |n+1 \rangle, \\
|\psi^{(0)}_{b} \rangle= C'_1 |n \rangle+ C'_2 |n+1 \rangle,
\end{gather}
where normalized coefficients  are
\begin{gather}
|C_1|^2= \frac12\left(1- \frac{\delta}{\Delta E} \right), \quad
|C_2|^2= \frac12\left(1+ \frac{\delta}{\Delta E}
\right), \\
|C'_1|^2= \frac12\left(1+ \frac{\delta}{\Delta E} \right), \quad
|C'_2|^2= \frac12\left(1- \frac{\delta}{\Delta E} \right)
\end{gather}
The corresponding energy levels are ($E_a < E_b$)
\begin{equation}
E_{a}= \varepsilon_{n}- \frac12(\Delta E+ \delta), \quad E_b=
\varepsilon_n+ \frac12(\Delta E- \delta),
\end{equation}
The value of $E_{a}$ gives the energy of the ground state of the
Hamiltonian $\widetilde h_0$. Here $\Delta E= \sqrt{\delta^2+ 4t^2
|\psi|^2 (n+1)}$ is the splitting between the two states, $E_b-
E_a= \Delta E \ge |\delta|$, and $\delta= \mu- n= \varepsilon_n-
\varepsilon_{n+1}$, $-1/2 < \delta < 1/2$. In the Mott phase
$|\psi|= 0$, the corresponding ground state wave function is
either $|\psi^{(0)}_a \rangle= |n+1 \rangle$ for positive $\delta
>0$ ($\Delta E= \delta$), or $|\psi^{(0)}_a \rangle= |n \rangle$ for negative $\delta <0$
($\Delta E= -\delta$).

At integer values of $\mu$ ($\mu= n$, $\delta= 0$) the splitting
is proportional to the magnitude of the order parameter, $\Delta
E= 2t|\psi| \sqrt{n+1}$. To this non-analytical dependence of
ground state energy on $\psi$ at integer values of $\mu$, $E_{gs}=
t|\psi|^2+ E_a= -n(n+1)/2+ t|\psi|^2- t|\psi| \sqrt{n+1}$, was
pointed out in Ref. \cite{Stoof01}.

According to the standard perturbation theory the first order
correction to the ground state wave function is
\begin{equation}
|\psi_a^{(1)} \rangle= \sum \limits_{k \ne n, n+1} \frac{\langle
k|h''_t|\psi_a^{(0)} \rangle}{E_a- \varepsilon_k} |k \rangle
\end{equation}
Of three terms of $h''_t$, Eq. \eqref{ht}, only term
$P^{\perp}h_tP$ gives a contribution into matrix element $\langle
k|h''_t|\psi_a^{(0)} \rangle$. As a result the perturbed wave
function is
\begin{multline} \label{wf2}
|\psi_a \rangle= |\psi_a^{(0)} \rangle + |\psi_a^{(1)} \rangle= \\
= C_1 |n \rangle+ C_2 |n+1 \rangle+ \\ + C_1
\frac{(-t)\psi^*\sqrt{n}}{E_a- \varepsilon_{n-1}} |n-1 \rangle+
C_2 \frac{(-t)\psi\sqrt{n+2}}{E_a- \varepsilon_{n+2}} |n+2 \rangle
\end{multline}
Comparing Eqs. \eqref{wf1} and \eqref{wf2} one can clearly see the
advantage of this approach, namely the coefficients of the
decomposition of $|\psi_a \rangle$ have regular behavior as
functions of $\mu$.

Self-consistency equation for the order parameter, $\psi= \langle
\psi_a |b |\psi_a \rangle$, is reduced to equation
\begin{multline} \label{bndr2}
\psi= t \psi \chi_a, \quad \chi_a= \frac{1}{\Delta E} \left[
(n+1)+ n \frac{\Delta E-
\delta}{2+ 3\delta+ \Delta E}+ \right. \\
+\left. (n+ 2) \frac{\Delta E+ \delta}{2- 3\delta+ \Delta E}
\right]
\end{multline}
Close to the critical boundary, $|\psi|^2 \simeq 0$, Eq.
\eqref{bndr2} can be rewritten as
\begin{gather} \label{bndr3}
\psi= \left\{ \begin{aligned}
&t\psi \chi_0(\mu, n+1), \quad & \delta > 0, \\
&t\psi \chi_0(\mu, n), \quad   & \delta < 0
\end{aligned}
\right.
\end{gather}
Eq. \eqref{bndr3} gives the lower part of the lobe $\mu(t,n+1)$ at
$n \leq \mu \leq n+ 1/2$ and the upper part of the lobe $\mu(t,n)$
at $n- 1/2 \leq \mu \leq n$. Combining parts from different $n$
results in standard mean-field lobes as described by Eq.
\eqref{bndr1}.

Another way to obtain the position of the phase boundary in the
plane $(\mu, t)$ is to consider a correction to the ground state
energy. It is a sum of three terms, $\Delta E_{gs}= t |\psi|^2+
(E_a- E_{gs}^{(0)})+ \Delta E^{(2)}_{gs}$, $E_{gs}^{(0)}= min
(\varepsilon_{n}, \varepsilon_{n+1}))$, originating from: (i) the
mean-field treatment of the hopping term, (ii) a correction due to
the formation of the state $|\psi_a^{(0)} \rangle$, $E_a-
E_{gs}^{(0)} \simeq -t^2 |\psi|^2(n+1)/ \Delta E$, and (iii) the
second order correction due to the perturbation $h''_t$,
\begin{multline}
\Delta E^{(2)}_{gs}= t^2 |\psi|^2 \left[ \frac{|C_1|^2n}{E_a-
\varepsilon_{n-1}}+ \frac{|C_2|^2(n+2)}{E_a- \varepsilon_{n+2}}
\right]= \\
= -t^2 |\psi|^2 \frac{1}{\Delta E} \left[ n \frac{(\Delta E -
\delta)}{2+ 3\delta+ \Delta E}+ (n+2)\frac{(\Delta E + \delta)}{2-
3\delta+ \Delta E} \right]
\end{multline}
Gathering all three terms one obtains the correction to the ground
state energy, $\Delta E_{gs}= t |\psi|^2(1- t\chi_a)$, where
$\chi_a$ is given  by Eq. \eqref{bndr2}. At the phase boundary,
$\Delta E= |\delta|$, and $\chi_a= \chi_0(\mu, n+1)$ at $\delta
> 0$, or $\chi_a= \chi_0(\mu, n)$ at $\delta
> 0$. One recovers the upper and lower parts of the corresponding
lobes.

\begin{figure}
\includegraphics[width=8.9cm]{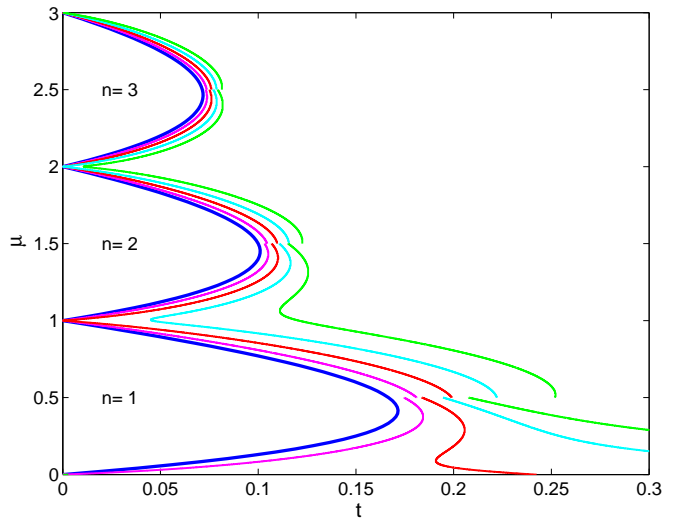}
\caption{\label{Fig1} Mean- field phase diagram for the BH model
\eqref{ham}. Thick lines indicate the Mott insulating lobes for
$n= 1, 2$ and 3. Thin lines are contours of constant value of the
order parameter, corresponding to $|\psi|^2= 0.2, 0.4, 0.6$ and
0.8.}
\end{figure}

From Eq. \eqref{bndr2} one enables also obtain some qualitative
information about the behavior of order parameter in the
superfluid phase. Namely, one can recover the dependence of the
order parameter $|\psi|^2$ on parameters $\mu$ and $t$. This
dependence is shown in Fig.~ \ref{Fig1}. Mott insulating lobes
(thick lines) coincide with the prediction of standard
perturbation approach, Eq. \eqref{bndr1}. Thin lines outside the
insulating lobes correspond to a few contours of $|\psi|^2= const$
inside the superfluid phase. The contour plot is discontinuous at
points corresponding to half-integer values of $\mu$. This can be
expected from our restriction of the Hamiltonian $h_t$ to the
two-dimensional subspace ${\cal P}^2_n$. By going from ${\cal
P}^2_n$ to another ${\cal P}^2_{n+1}$ matrix elements of the
restricted perturbation $h'_t$ jump to new values. This artificial
feature can be circumvented if to use the bigger vector space of
number states $\{ |n \rangle \} , n= 0, 1, \cdots, n_{max}$. Then
it will be equivalent to the approach of Refs. \cite{Jaksch98,
Trefzger08}.

In summary, we presented an alternative way to estimate
perturbation corrections to the ground state wave function of the
mean field BH Hamiltonian \eqref{ham}. This approach has an
advantage that the corrections to $|\psi_{gs} \rangle$ and
$E_{gs}$ do not have singularities at integer values of the
chemical potential. It enables, as well, to describe the behavior
of $|\psi|$ also in the superfluid phase. However, the restriction
of the perturbation onto the two-dimensional subspace ${\cal
P}^2_n$ introduces the discontinuity in the behavior of order
parameter $|\psi|$ at half-integer values of $\mu$ in the
superfluid phase.

The fact that one obtains the same equation for critical boundary,
$t= 1/\chi_0(\mu, n)$, using degenerate and non-degenerate
(standard) perturbation expansion tells that this equation is the
generic property of the on-site mean-field approximation and is
not connected with the way how one take into account degeneracy of
spectrum $\varepsilon_n$ at integers values of $\mu$.

The work was supported by RFBR-NSFC [Grant No. 13-02-91177 and No.
13-02-00909].

\end{document}